# Aligning Multiple Protein Structures using Biochemical and Biophysical Properties

Paul Shealy and Homayoun Valafar

Department of Computer Science

University of South Carolina

Columbia, SC 29208

**Abstract** - Aligning multiple protein structures can yield valuable information about structural similarities among related proteins, as well as provide insight into evolutionary relationships between proteins in a family. We have developed an algorithm (msTALI) for aligning multiple protein structures using biochemical and biophysical properties, including torsion angles, secondary structure, hydrophobicity, and surface accessibility. The algorithm is a progressive alignment algorithm motivated by popular techniques from multiple sequence alignment. It has demonstrated success in aligning the major structural regions of a set of protein from the s/r kinase family. The algorithm was also successful at aligning functional residues of these proteins. In addition, the algorithm was also successful in aligning seven members of the acyl carrier protein family, including both experimentally derived as well as computationally modeled structures.

## 1 Introduction

Comparisons of protein structures can yield valuable information about biologically relevant similarities between related structures. Identification of common structural motifs between two proteins can provide valuable information about their evolutionary relationship and yield insight into structural components required for the proteins to function. A structure alignment can help identify the function for a novel protein. Furthermore, structural comparison algorithms are an important validation step for approaches to protein folding, such as *ab initio* methods or threading algorithms.

While a pairwise comparison between two structures is useful, a simultaneous analysis of multiple structures can be far more informative. With more than 50,000 protein structures in the PDB [1] as of 2009, they can be classified into a hierarchy by structure and function, such as the manually curated database SCOP [2]. This hierarchy provides a starting point for a variety of investigations, such as determining the function of a new protein with a known structure or isolating the functional residues from a set of related structures.

Comparing protein structures is an inherently difficult task. One problem is that is no single accepted definition of structural similarity. Many structure alignment algorithms consider the protein to be a single rigid entity and use geometric properties to define similarity. Some algorithms rely on interatomic distance maps [3; 4]. Others use the distance and orientation of secondary structure elements [5] or distance comparisons between heptapeptide fragments [6], Recently, algorithms on multiple structure alignment have been developed, using heptapeptide fragments [7], backbone RMSD of protein fragments of varying size [8], interatomic distance maps with a Monte Carlo search [9], or secondary structure elements [10].

Here we present an algorithm called msTALI (multistructure torsion angle alignment) which utilizes local structural information to create alignments of multiple protein structures. msTALI is inspired by and extends previous work on pairwise structure alignment using torsion angles [11]. It computes alignments using local structural motifs derived from torsion angles and biochemical properties. Proteins with very distant evolutionary relationships are more likely to exhibit local structural similarities. In addition, alignments using torsion angles are computationally efficient when computing the optimal solution.

## 2 Background and Method

### 2.1 Pairwise Structure Alignment

Pairwise alignment of protein structures is a core component of the multiple structure alignment mode of msTALI. This algorithm is a straightforward extension of sequence alignment algorithms to structures. It treats each protein as a series of residues and applies a generalized Needleman-Wunsch [12] algorithm to find the best global alignment. A comparison between two residues is based on their biochemical and biophysical properties. One core property is the residue's torsion angles. Treating a protein as a polymer of peptide planes, the relationship between two consecutive peptide planes can be described by two torsion angles, φ and ψ. The atomic positions of a protein's backbone can be almost entirely determined by a complete set of torsion angles. Representing a protein in this form provides a compact, concise representation of the protein backbone.

While all torsion angles are possible for a residue, many values yield an atomic structure that is highly unfavorable due to steric hindrance and London dispersion forces and therefore are rarely observed. Other values yield secondary structures, α-helices and β-sheets, which are locally stable. The space of all possible torsion angles and their frequency of observance is called a *Ramachandran space* [13]. Ramachandran space has regions of high probability that correspond to secondary structures, with much of the remainder corresponding to angles of low probability. To compare two sets of torsion angles, the difference in probabilities is used. Because separate secondary structure regions may have similar probabilities, a separate penalty is imposed for transitions across a secondary structure boundary. The secondary structure scoring function is then

$$S_T(r_a, r_b) = |R(r_a) - R(r_b)| + Err(r_a, r_b) \quad (1)$$

where $S_T$ is the torsion angle scoring function, $r_a$ and $r_b$ are the residues to be scored, $R$ is the Ramachandran likelihood function, and $Err$ is the secondary structure penalty function.

Additional properties used to compare two residues are hydrophobicity and surface accessibility. The hydropobicity scale used is the Kyte-Doolittle scale [14]. Hydrophobicities for two residues are compared by computing the difference between the hydrophobicity values for the residues. Surface accessibility is computed using DSSP [15], and the two values are compared as the difference in values. Finally, sequence information can be incorporated if desired. The individual penalty functions are

$$S_H(r_a, r_b) = |H(r_a) - H(r_b)| \quad (2)$$

$$S_A(r_a, r_b) = |A(r_a) - A(r_b)| \quad (3)$$

$$S_S(r_a, r_b) = M(r_a, r_b) \quad (4)$$

Where $S_H$, $S_A$, and $S_S$ are scores for the hydrophobicity, surface accessibility, and sequence scores, respectively; $H(r)$ is the hydrophobicity of residue $r$, $A(r)$ is the surface acccessibility, and $M$ is the scoring matrix.

The final scoring function for two residues is

$$S(r_a, r_b) = \begin{aligned} & w_T \cdot S_T(r_a, r_b) \\ & + w_H \cdot S_H(r_a, r_b) \\ & + w_A \cdot S_A(r_a, r_b) \\ & + w_S \cdot S_S(r_a, r_b) \end{aligned} \quad (5)$$

Where $w_T$, $w_H$, $w_A$, and $w_S$ are the weights for the individual components.

Gaps are implemented using a standard affine gap penalty. For a gap of *n* residues, with a gap open penalty $G_O$ and gap extension penalty $G_E$, the total penalty is

$$G(n) = G_O + n \cdot G_E \quad (6)$$

## 2.2 Multiple Structure Alignment

The pairwise alignment algorithm can be extended to multiple structure alignment in a straightforward manner. This algorithm is a progressive alignment algorithm inspired by ClustalW [16]. This algorithm is based on the observation that the easiest alignments to compute are between structures that are the most similar. The algorithm to compute the multiple structure alignment between a set *S* of protein structures is as follows:

1. Compute the pairwise distances between any two structures in *S*.

2. Compute a guide tree for *S*, using the pairwise distances from step 1.

3. Progressively align the structures of *S* according to the guide tree, working from the leaves to the root.

This algorithm utilizes a structure *profile*, which is a set of aligned structures. In step 1, pairwise distances are computed between all pairs of structures in *S*, using the pairwise alignment method described earlier. These distances are used to compute a guide tree in step 2, using the neighbor-joining algorithm

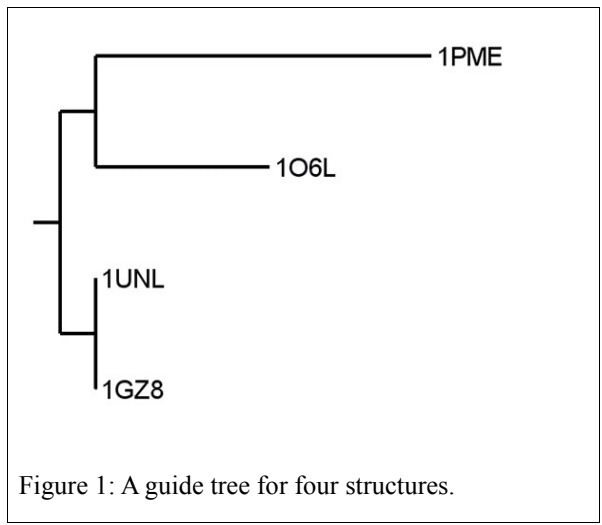

Figure 1: A guide tree for four structures.

[17]. An example is shown in Figure 1. Once the guide tree is constructed, each leaf node is associated with an alignment profile containing only that node's structure, so each structure belongs to a separate profile. Each profile is weighted according to its distance from the root in the guide tree [16]. Weighting reduces the impact of several highly similar structures on the final alignment.

In step 3, two leaf nodes that share a common parent node (i.e., siblings in the tree) are selected, and their associated profiles are aligned. The aligned profiles are combined into a single profile, and the parent node joining the leaves is replaced by a single node associated with the combined profile. This is repeated until only a single profile remains, which is the full alignment. In Figure 1, the structures are aligned as follows: 1PME versus 1O6L, 1UNL versus 1GZ8, and 1UNL/1GZ8 versus 1PME/1O6L.

To align two profiles, the same core alignment algorithm in step 1 is used. To score a residue-to-residue match between a position in profiles *p1* and *p2*, each residue from *p1* is compared to each residue from *p2* using the scoring method described earlier for two residues, the score is multiplied by the weights from the two structures, and the average weighted score is computed. The score for a residue versus a gap is zero. Because all scoring metrics used are scaled to be non-negative, a residue versus gap score is the lowest possible value.

The structures in a profile remain fixed with respect to one another during the rest of the algorithm. Thus, the residues at each position do not shift, and once a gap is introduced in a profile, it remains fixed for the remainder of the alignment. When introducing a gap in a profile, the gap is introduced in all structures in the profile.

Scoring an affine gap requires two parameters: a gap open penalty and a gap extension penalty, referred to as *GO* and *GE*, respectively. These are specified by the user. However, both parameters are modified by the algorithm as follows:

1. If there are existing gaps at the new gap location, the penalties are adjusted as follows, where $ng$ is the number of gaps at the location and $ns$ is the number of structures. In this case, no other rules apply.

$$GO \leftarrow 0.3 \cdot GO - 0.3 \cdot ng/ns \cdot GO$$

$$GE \leftarrow 0.5 \cdot GE$$

2. If there is a loop at the new gap location, the penalties are adjusted as follows, where $nl$ is the number of profiles with a loop at the location and $ns$ is the number of structures.

$$GO \leftarrow GO - 0.75 \cdot GO \cdot nl/ns$$

$$GE \leftarrow GE - 0.5 \cdot GE \cdot nl/ns$$

Often, the initial alignment will produce reasonable results that can be improved by *realignment*. Realigning a profile involves repeatedly extracting a single structure from it, then aligning it to the remaining profile. Here we realign each structure once.

Using a guide tree to inform the order in which to align the structures causes the most similar structures to be aligned first, leading to subtle features in the alignment being identified early. Structures are weighted to reduce the influence of multiple highly related structures on the final alignment.

## 3 Results and Discussion

We applied msTALI to two sets of proteins. One, s/r kinases, is a set of four highly conserved structures with a low sequence conservation of 17% across the set. The second set of proteins, acyl carrier proteins, contains six structures of identical function but with a number of small structural variations between members.

For all experiments here, the weights used were

$$w_T = 0.5$$
$$w_H = 0.2$$
$$w_A = 0.3$$
$$w_S = 0.0$$

where the contributions to the total score are 50% torsion angles, 20% hydrophobicity, 30% surface accessibility. The sequence component is unused. Because some key residues are conserved across all proteins being aligned, this provides a method for validation of the algorithm.

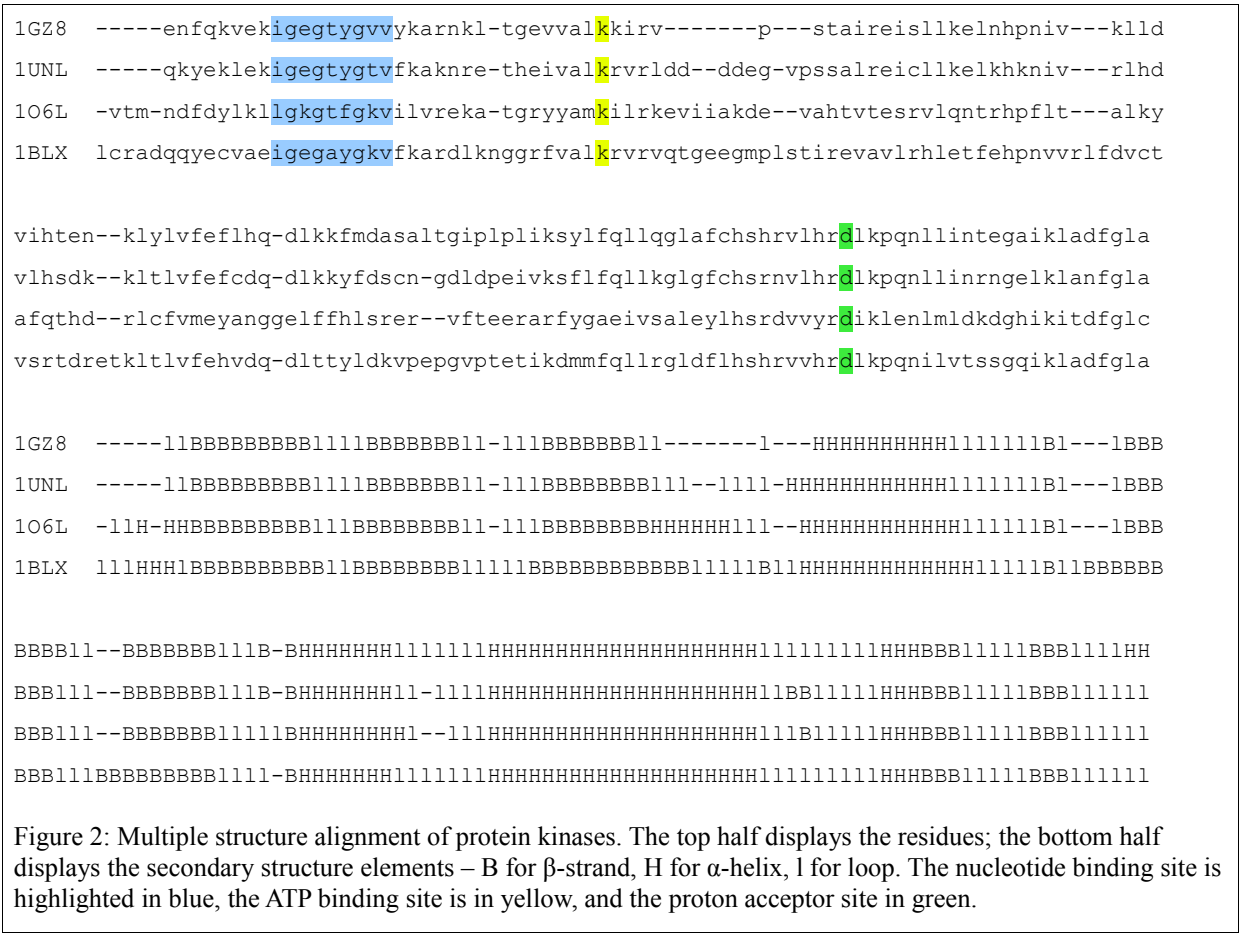

Figure 2: Multiple structure alignment of protein kinases. The top half displays the residues; the bottom half displays the secondary structure elements – B for β-strand, H for α-helix, l for loop. The nucleotide binding site is highlighted in blue, the ATP binding site is in yellow, and the proton acceptor site in green.

## 3.1 Protein S/R Kinases

We aligned four members of the protein kinase catalytic subunit family as indicated in SCOP: 1GZ8, 1UNL, 1O6L, and 1BLX. All are serine/threonine kinases from *homo sapiens*. The entire chain indicated in SCOP was aligned, although the s/r domain is the first ~50% of each chain. Here we only display the alignment of the s/r kinase domain for brevity.

These structures were simultaneously aligned using msTALI. No sequence information was used to create the alignment. Because many of the structures have a high sequence identity to other structures in the alignment, sequence information provides a valuable, independent method for judging the quality of the alignment.

The full alignment is displayed in Figure 2. The score is an indication of the degree of conservation at a position, ranging from a low of 0 to a high of 9.

The majority of the secondary structure regions are conserved across all structures; in all cases, msTALI has aligned these regions together. Furthermore, for many of these conserved secondary structures of equal length, msTALI has aligned the ends of the structures. Where there is ambiguity about the precise alignment, such as the additional α-helices present in 1O6L, the hydrophobicity and surface accessibility components provide valuable information in determining the exact local alignment.

The functional regions of the protein s/r kinases are annotated in UniProtKB [18]. These sites are highlighted in Figure 2. These sites were all aligned correctly by msTALI. Furthermore, these sites are all scored highly by msTALI, indicating well-conserved sites.

## 3.2 Acyl Carrier Proteins

The acyl carrier protein (ACP) family is involved in fatty acid synthesis, linking intermediates during synthesis via a thioester linkage. Here we have aligned a set of acyl carrier proteins from varying sources. Three are crystal structures: 2FAC, 1L0I, and 2EHS. Two are NMR structures: 2JQ4 and 1ACP. One,

```
2FAC   tieervkkiigeqlgv--kqeevtnnasfvedlgadsldtvelvmaleeefdteipdeeaek--ittvqaaidyin---g-hq-
1L0I   tieervkkiigeqlgv--kqeevtnnasfvedlgadsldtvelvmaleeefdteipdeeaek--mttvqaaidyin---g-hq-
2EHS   -leervkeiiaeqlgv--ekekitpeakfvedlgadsldvvelimafeeefgieipdedaek--iqtvgdvinylk---e-k--
2JQ4   --natireilakfgqlptpvdtiadeadl-yaaglssfasvqlmlgieeafdiefpdnllnrksfasikaiedtvklildgkea
1ACP   tieervkkiigeqlgv—kqeevtnnasfvedlgadsldtvelvmaleeefdteipdeeaek--ittvqaaidyin---g-hq-
AcpXL  atfdkvadiiaetsei--dratitpeshtiddlgidsldfldivfaidkefgikiplekwtq---e-----vn----------

2FAC   lHHHHHHHHHHHHll--lHHHlllllBlllllllllHHHHHHHHHHHHHHlllllllHHHHl--llBHHHHHHHH---H-Hl-
1L0I   lHHHHHHHHHHHHll--lHHHlllllBlllllllllHHHHHHHHHHHHHHlllllllHHHHll--llBHHHHHHHH---H-ll-
2EHS   -HHHHHHHHHHHHll--lHHHlllllBlllllllllHHHHHHHHHHHHHHlllllllHHHHHl--llBHHHHHHHH---H-H--
2JQ4   --HHHHHHHHHllllllllHHHlllllllH-HHHlllHHHHHHHHHHHHHHHlllllHHHHllHHHHlHHHHHHHHHHHHlHHH
1ACP   llHHHHHHHHHHHlll--lllllllllllllllllHHHHHHHHHHHHHHHlllllllHHHHll--llllHHHHHHHH---H-Hl-
AcpXL  lHHHHHHHHHHHHll—lHHHlllllBlllllllllHHHHHHHHHHHHHHlllllHHHHll---l-----lB----------

Score  45889988898888890087899889878688888899889999889898878998898888006766667866600060530
```

Figure 3: Secondary structure of the residues of acyl carrier proteins. H denotes an α-helix (of any type), B denotes a β-bridge or β-sheet, and l denotes a turn, bend, or loop.

AcpXL, is computationally modeled using I-TASSER [19]. These structures were simultaneously aligned using msTALI.

The full alignment of all acyl carrier proteins structures is shown in Figure 3. The three crystal structures are aligned perfectly with respect to one another. One NMR structure, 1ACP, has high sequence similarity to the crystal structures. It is clearly aligned well from its sequence identity and secondary structure similarity to the crystal structures. The second NMR structure, 2JQ4, has much lower sequence identity, and so its alignment is more difficult to evaluate. However, the conserved secondary structure elements are aligned well with respect to the crystal structure 2FAC. The conserved secondary structure elements – the first, second, fourth, and fifth α-helices of 2JQ4, are aligned to their corresponding helices from 2FAC. The alignment of the third α-helix of 2JQ4 to loop regions from other structures is reasonable, given that the surrounding helices are precisely aligned and only a single gap was inserted in this region. Finally, 2JQ4 has three α-helices at its C-terminus, whereas the other structures only have one. TALI has aligned the second

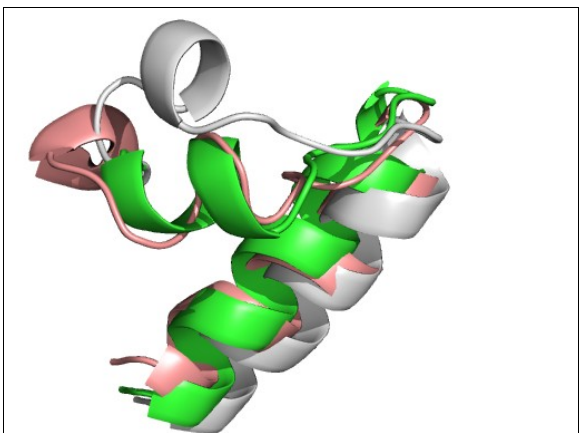

Figure 4: Alignment of the crystal structure 2EHS, in green, the NMR structures 2JQ4, in magenta, and 1ACP, in silver.

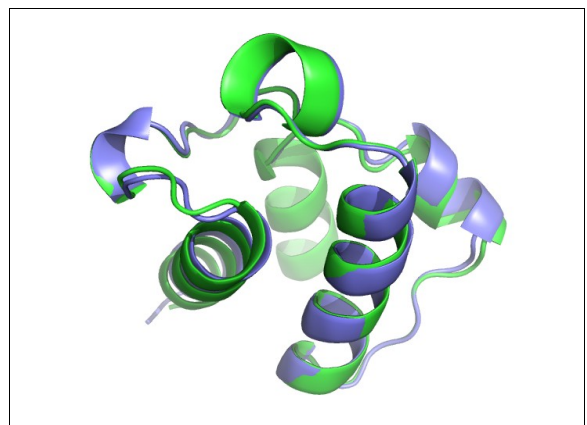

Figure 5: Alignment of the crystal structure 2EHS, in green, to the computational structure AcpXL, in purple.

helix from 2JQ4 to the single helix from the other proteins.

The computational structure, AcpXL, has a high sequence identity to the crystal structures, so the alignment is easy to assess. In addition, the secondary structure elements are identical. From these items, we see that these structures are precisely aligned to their crystal structure counterparts. The only questionable part is the tail of AcpXL, which appears to be shifted to the right by six residues.

The (rigid) protein structures were aligned in MolMol [20] using the major conserved region from each protein, as indicated by msTALI. This region is annotated in Figure 3.

A full alignment of this set of structures provides valuable insight into this protein family. When considering only the three crystal structures, the structures are highly similar, with pairwise backbone RMSDs ranging from 0.39 to 0.67. Incorporating the NMR structures as well provides insight into the regions of similarity and difference between the crystal and NMR structures. Figure 4 displays a partial region from this alignment, isolating the two NMR structures and a single crystal structure. The α-helices on the right are aligned well, while the helical regions in the top left show a clear divergence. These regions of divergence could be highly dynamical regions or areas of varying functionality.

Finally, displaying the multiple structure alignment between the crystal structure 2EHS and the computationally modeled structure AcpXL provides valuable information on the conservation of core structural and functional residues in the modeled structure. This is illustrated in Figure 5.

## 4 Conclusion

We have presented an algorithm for aligning multiple protein structures. It is computationally efficient, with a computational complexity on the same order as multiple sequence alignment with ClustalW. It provides both pairwise and multiple structure alignments using several germane biochemical and biophysical properties.

An alignment between multiple protein structures provides a wealth of information about a protein family. For example, aligning proteins that contain a common active site could provide a method to identify the key residues in the active site. It could also identify the structural elements required to position the atoms in the active site correctly. Aligning structurally related proteins can also elucidate their structural differences, which may provide insight into the structural components required to perform particular functions.

msTALI is currently implemented in MATLAB®, a flexible development environment centered around matrices. We plan to re-implement it in C++ for speed and general-purpose use, and also to make a web version available. Furthermore, msTALI could be beneficial in a number of other tools from our lab, such as PDPA.

**Acknowledgements**

Thanks to the Rothberg Fellows program in the Department of Computer Science and Engineering at the University of South Carolina for their support to PGS.